# Self-powered InAs nanowire detector arrays for extended-SWIR spectrometry at room temperature

Yang Yu[a], Wei Wen Wong[a], Zhe Li[a], Dawei Liu[a], Jinyuan Chen[b], Seyed Saleh Mousavi Khaleghi[b], Yue Bian[a], Kosala Dhanawansha[a], Li Li[a], Hongwei Liu[c], Xiaoxue Xu[d], Monica S. Allen[e], Jeffery W. Allen[e], Hark Hoe Tan[a], Chennupati Jagadish[a], Kenneth Crozier[b], Ziyuan Li[a,*], Chaohao Chen[d,f*], and Lan Fu[a,*]

[a] *Australian Research Council Centre of Excellence for Transformative Meta-Optical Systems, Department of Electronic Materials Engineering, Research School of Physics, The Australian National University, Canberra, ACT 2600, Australia*

[b] *Australian Research Council (ARC) Centre of Excellence for Transformative Meta-Optical Systems (TMOS), University of Melbourne, Victoria, Australia*

[c] *The Australian Centre for Microscopy and Microanalysis, The University of Sydney, Sydney, New South Wales, 2006, Australia*

[d] *School of Biomedical Engineering, Faculty of Engineering and Information Technology, University of Technology Sydney, NSW 2007, Australia*

[e] *Air Force Research Laboratory, Air Warfare Directorate, Eglin AFB, Florida 32542, USA*

[f] *Australian Research Council Centre of Excellence for Transformative Meta-Optical Systems, The University of Technology Sydney, Ultimo, NSW 2007, Australia*

*Corresponding emails: ziyuan.li@anu.edu.au, chaohao.chen@uts.edu.au, lan.fu@anu.edu.au*

# Abstract

Spectral sensing in the extended shortwave infrared (e-SWIR) is important for molecular analysis, infrared imaging, and machine vision, motivating the development of compact spectrometers for broader applications. However, conventional commercial off the shelf spectrometers in this wavelength region are expensive and bulky due to their reliance on external dispersive optics/filters and/or cryogenic accessories. Other emerging computational spectrometry is based on Si and InGaAs photodetectors that remain focused on the visible and near-infrared, with few detector platforms operating in the e-SWIR regime that simultaneously provides broadband sensitivity, low-noise room-temperature operation and diverse spectral signatures for accurate identification and reconstruction. Here, we report a room-temperature e-SWIR computational spectrometer based on InAs/InP core-shell nanowire photodetector arrays with geometry-encoded spectral responses. The detectors exhibit self-powered broadband photoresponse across the 1–3 μm range, with responsivity up to 0.215 A $W^{-1}$, detectivity up to $1.6 \times 10^{9}$ cm·$Hz^{1/2}$ $W^{-1}$, and microsecond response times. The excellent detector performance is leveraged to demonstrate filter-free spectral reconstruction using a compact multipixel photodetector array device. This enables high-accuracy molecular absorption spectrum reconstruction and hyperspectral imaging. Our results indicate that InAs nanowire arrays are a promising platform for compact computational spectrometry and imaging in the e-SWIR at room-temperature.

# Introduction

Spectral sensing in the extended shortwave infrared (e-SWIR, 1.7-3 μm) is important for molecular identification, chemical analysis, and infrared imaging.[1] This spectral range contains molecular overtone and combination absorption bands that are weak or inaccessible in the visible, making it particularly valuable for compact sensing and imaging technologies. Accessing these spectral signatures requires spectrometers capable of resolving wavelength-dependent optical signals across a broad infrared band. Conventional spectrometers operating in the e-SWIR have high size, weight and cost because they typically rely on bulky dispersive or interferometric optics and cooling-assisted detectors. This limits miniaturization, integration, and practical application for on-chip sensing and imaging systems on small frame systems. In contrast, computational spectrometry offers a compact alternative by reconstructing incident spectra from wavelength-dependent detector responses.[2-12] Despite the rapid progress in this field, most reported implementations remain focused on the visible and near-infrared, typically below ~1.7 μm, relying on well-established Si[13] and lattice-matched InGaAs detector[14,15] technologies for high signal quality. Consequently, room-temperature demonstrations at longer wavelengths remain comparatively scarce, leaving the crucial molecular and heat signatures beyond 2 μm largely unexploited.[16]

Some recent efforts have looked at extending computational spectrometry into the e-SWIR and adjacent mid-infrared regions using spectrally encoded detector arrays or electrically tunable single detectors. Stack-effect tunable black phosphorus (BP) photodetectors were demonstrated for a wide operating spectral range of 2-9 μm, but require cryogenic operating temperatures and long-term stability remains an issue which limits their application and scalability.[6] Room-temperature demonstrations, on the other hand, have been severely constrained by low detector performance and noise. One recent example demonstrated a room-temperature spectrometer covering 2.2 to 3.8 μm using 20 Al-grating detectors fabricated on doped Si substrates and a Gaussian sparse reconstruction model, but with a low detector responsivity of 0.17 mA $W^{-1}$.[17] Another was based on a self-powered BP/$MoS_2$ heterojunction enabled room-temperature spectral reconstruction from 1.7 to 3.6 μm, but again exhibited low responsivity (~ mA $W^{-1}$ level) and detectivity ($7 \times 10^7$ cm·$Hz^{1/2}$·$W^{-1}$), and required encapsulation due to poor material stability in ambient environment.[7] While these studies show the feasibility of computational spectrometry beyond 1.7 μm, there are significant challenges that remain in detector performance (responsivity vs operating temperature), and scalability (manufacturing and packaging constraints).

Traditional narrow-bandgap III-V semiconductors such as InAs[18,19] are attractive for e-SWIR photodetection because of their intrinsic long wavelength sensitivity and high carrier mobility. Conventional planar InAs photodetectors suffer from high dark currents associated with high intrinsic carrier concentration that degrades room-temperature performance. Nanowire arrays have been shown to reduce the active material volume, to enhance absorption through geometry tuned optical resonances,[20] and to form radial p-n heterojunction for efficient short-distance carrier collection.[21,22] Moreover, their resonant absorption can be systematically tuned using the nanowire diameter and array pitch, providing a direct route to tailoring spectral response functions for computational spectrometry using geometry.[11,23] Previous InAs nanowire array photodetectors predominantly employed axial junction between nanowires and substrate. Devices grown and fabricated on InAs substrate achieved broadband response extending to 3.5 μm, but the narrow bandgap substrate contributed substantially to high dark current.[24] Heteroepitaxial InAs nanowire growth on wider bandgap InP and Si substrates performed better, but diameters were limited to ~100 nm leading to small optical cross-sections with weak long-wavelength absorption and limited infrared photoresponse.[25,26] Additionally, in these nanowires, extending the diameter range is challenging due to lattice-mismatched nucleation and growth instability.[27]

Here, we report the development of uniform, highly crystalline InAs/InP core-shell multipixel nanowire arrays on InP substrates whose spectral response can be tuned using diameter for demonstration of a self-powered room-temperature e-SWIR computational spectrometer. We employ a three-stage site-selective epitaxial growth strategy, with an InAsP double-buffer layer to accommodate lattice mismatch between the subsequent InAs nanowire growth on the InP substrate, and a thin InP shell for surface passivation, to establish a high quality nanowire material platform for e-SWIR photodetection. Further optimization and engineering of InAs/InP/ITO radial carrier-selective junction yielded efficient charge separation and dark-current suppression, high performance, room-temperature InAs nanowire arrays for e-SWIR photodetection. These fabricated arrays demonstrated tunable, broadband responses across 1 - 3 μm, a peak responsivity of 0.215 A $W^{-1}$, detectivity of $1.6 \times 10^9$ cm·$Hz^{1/2}$·$W^{-1}$, microsecond response times and single-pixel imaging under zero-bias operation. The nanowire geometry, i.e., diameter and pitch, were carefully tuned to fabricate a compact multipixel nanowire array spectrometer chip that enabled filter-free spectral reconstruction with a mean peak-localization error of approximately 1.65 nm, achieving a high single-peak resolving capability of 17 nm, and a dual-peak resolving capability of 93 nm. This chips was for subsequently used to demonstrate molecular absorption reconstruction, hyperspectral imaging.

# Results and discussion

### InAs nanowire-array architecture for e-SWIR computational spectrometer

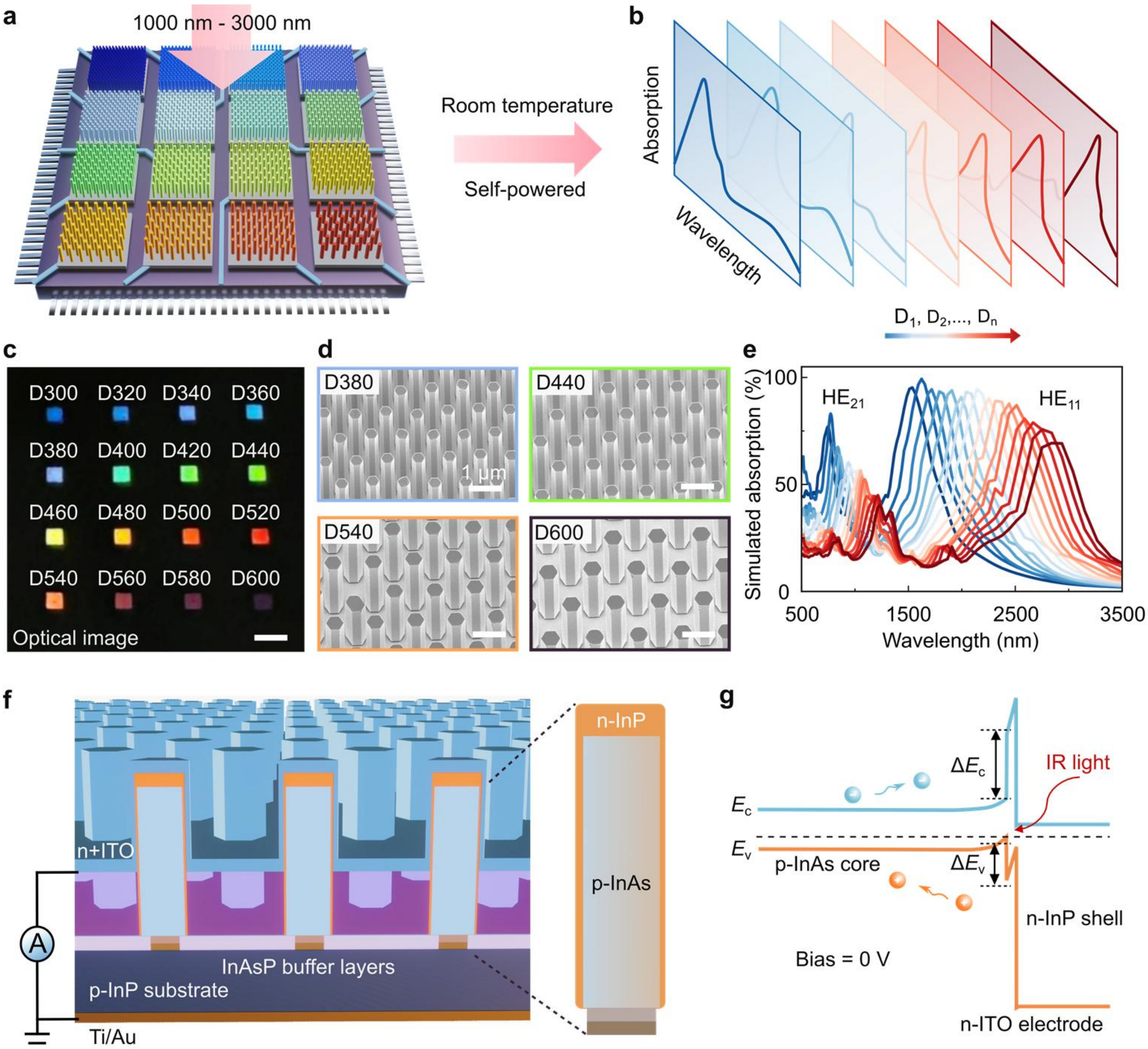

**Figure 1 | InAs-nanowire array architecture for room-temperature e-SWIR computational spectrometer. a**, Schematic illustration of the spectrometer architecture, comprising InAs nanowire arrays with progressively varied diameters and pitch grown on a single InP substrate. **b**, Conceptual illustration of the geometry-encoded spectral responses generated from detector arrays with engineered nanowire diameters and array parameters, forming the basis for spectral reconstruction. **c**, Optical images of the fabricated multipixel nanowire array library with nominal diameters ranging from 300 to 600 nm. Scale bar is 500

μm. **d**, Representative scanning electron microscopy (SEM) images of nanowire arrays with diameters of 380, 440, 540 and 600 nm. Scale bar is 1 μm. **e**, Simulated absorption spectra of corresponding InAs nanowire array with diameters ranging from 300 to 600 nm. **f**, Schematic of the nanowire photodetector device. The InAs/InP core-shell nanowire array is epitaxially grown on a p-InP substrate and contacted by $n^+$-Indium tin oxide (ITO) top and Ti/Au bottom electrodes. **g**, Calculated energy band diagram of InAs/InP radial p-n heterojunction with the n-ITO top contact under zero bias.

We designed a detector platform in which the spectral response is encoded directly using the geometry of InAs/InP nanowire arrays to realize room-temperature computational spectrometry in the e-SWIR. The overall concept is illustrated in **Figure 1a&b**. Arrays with different nanowire diameters and pitches were monolithically integrated on a single InP substrate. Each pixel provides a distinct wavelength-dependent response for spectral reconstruction. Unlike approaches that rely on external filters or optical components,[28,29] our strategy uses the nanowire arrays themselves as both the uncooled infrared absorbers and the spectral encoding elements, reducing weight and size by eliminating filtering and cryocooling components. This approach is particularly attractive for the e-SWIR, where broadband room-temperature absorption remains challenging in a compact form factor for scalable and diverse on-chip spectral encoding.

The optical image of as-grown InAs nanowire arrays is shown in **Fig. 1c**, with representative scanning electron microscopy (SEM) images of selected nanowire arrays presented in **Fig. 1d**. As the nominal nanowire diameter increases from 300 to 600 nm, and pitch size increases from 1200 to 2400 nm to ensure uniform nanowire growth. Correspondingly, the reflected colors of the arrays shift systematically from blue to red, indicating strong dependence of the optical response on geometry. Finite-difference time-domain (FDTD) simulations (details in **Methods**) in **Fig. 1e** further reveal that the nanowire arrays support two main resonant absorption peaks, which can be assigned to $HE_{21}$-like and $HE_{11}$-like hybrid modes (details in **Supplementary Fig. S8**).[11,23,30] As the nanowire diameter increases, both resonances redshift, with the $HE_{21}$-like resonance mainly located in the near-infrared region from ~750 - 1400 nm, and the $HE_{11}$-like resonance dominating the e-SWIR region from ~ 1500 - 2700 nm. The $HE_{11}$-like resonant maintains high peak absorption across the entire nanowire diameter range, producing distinguishable wavelength-dependent response profiles for computational spectral reconstruction.

The device architecture is illustrated in **Fig. 1f**. The p-type doped InAs nanowire arrays with different diameters were grown by selective-area epitaxy (SAE), through uniformly grown double InAsP buffer layers to mitigate effects of the substantial lattice mismatch between InAs and the wider-bandgap InP substrate. A conformal n-InP shell was subsequently grown *in situ* around the p-InAs core, simultaneously passivating surface states and establishing a radial p-n heterojunction. As illustrated by the energy band diagram (simulation details in **Methods**) in **Fig. 1g**, Fermi-level alignment induces band bending and a built-in radial electric field across the InAs/InP interface. Under illumination, this field separates photogenerated electrons and holes without an external bias, directing electrons towards the n-InP shell and electron-selective ITO contact while holes are collected through the p-InAs core. The InP shell also suppresses surface-state-mediated recombination and leakage, that plague narrow-bandgap InAs. Consequently, each nanowire array superpixel operates like a multifunctional element by simultaneously acting as a resonant infrared absorber, a geometry-programmed spectral encoding element, and a self-powered photodiode. This unique optical and electrical design provides broadband sensitivity and spectrally diverse response functions required for compact computational spectrometry.

**Uniform InAs/InP core-shell nanowire array growth**

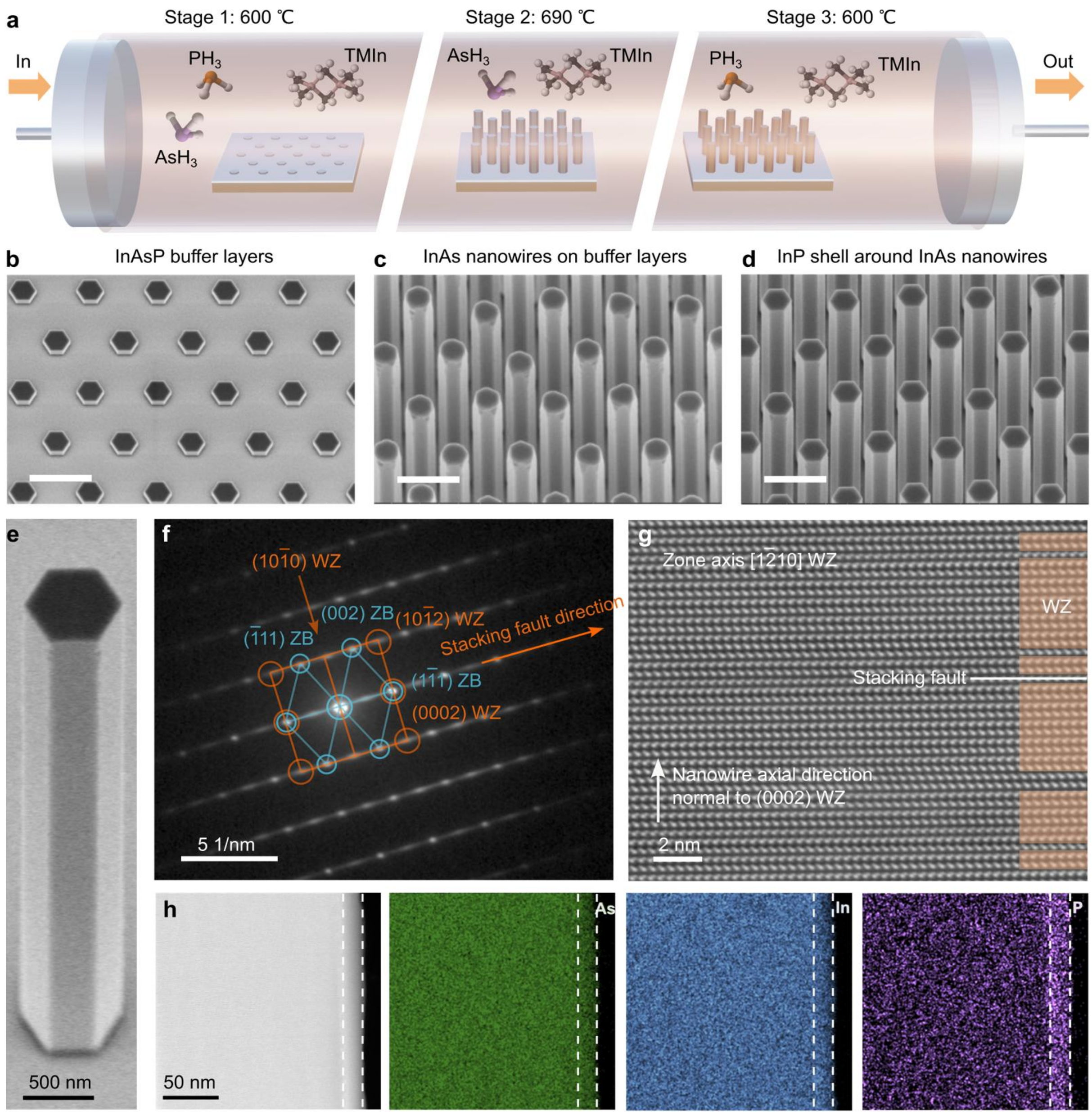

**Figure 2 | Growth and structural characterization of InAs/InP nanowire arrays grown on InP substrate. a**, Schematic diagram of the three-stage MOCVD growth sequence to realize the InAs nanowire arrays. A double InAsP buffer layer, InAs nanowires, and InP shells were grown sequentially at 600 °C, 690 °C, 600 °C, respectively. The precursor gases used for each step are indicated. **b – d**, 30°-tilted SEM images showing the growth evolution from the double InAsP buffer layers on the InP substrate (b) to the InAs nanowires grown on the buffer layers (c) and, finally, after the growth of a thin InP passivation shell layer (d). Scale bar is 1 μm. **e**, SEM image of a representative single nanowire with a diameter of 600 nm and height of 4 μm. **f**, Selected-area electron diffraction pattern of an InAs/InP nanowires, indexed to reveal the

dominant wurtzite crystal structure. **g**, Aberration-corrected HAADF-STEM image of the middle section of the nanowire, showing predominantly WZ atomic ordering. The image was acquired along the $[1\bar{2}10]$ WZ zone axis, and the arrow indicates the nanowire axial direction normal to the (0002) WZ basal planes. **h**, HAADF-STEM image of the core–shell interface at a nanowire sidewall and corresponding energy-dispersive X-ray spectroscopy elemental maps for As, In, and P, confirming the InAs/InP core-shell nanowire structure.

Demonstration of the geometry-encoded detector library requires uniform InAs/InP core-shell nanowire arrays spanning a broad and precisely controlled range of diameters and pitches for broad a highly sensitive spectral response, while maintaining effective surface passivation and low dark current at room-temperature to enable uncooled and low-noise operation. Growth of uniform, high-quality multiple InAs nanowire arrays with a wide range of large diameters on the same substrateis non-trivial. InAs nanowires can be grown on native substrates by SAE for strain-free growth and mask-tunable diameter but high dark currents associated with the substrate severely degrade overall photodetector performance.[24] Growth on wider-bandgap InP substrates provides an attractive alternative to reduce the substrate contribution to dark current with easy accessibility to commercial InP wafers. However, a common limitation among existing demonstrations is the lack of control over nanowire diameter. InAs nanowire diameters are typically limited to < 300 nm due to the large lattice mismatch with InP substrates.[25,27] Larger growth areas results in 3D nucleation and defective coalescence driven by strain relaxation,[31] hindering high-quality nanowire growth.

To overcome these limitations, we developed a three-stage site-selective nanowire growth process using SAE via metal-organic chemical vapor deposition (MOCVD), as illustrated in **Figure 2a** (see **Supplementary Fig. S1** for detailed growth sequence). The first stage was to grow an InAsP double-buffer layer to accommodate the lattice mismatch between the InAs nanowires and the InP substrate. The buffer layer stack, consisting of a P-rich and an As-rich InAsP layers, was grown to completely fill up the growth mask openings, as shown in Fig. 2b. The epitaxy sequence is followed by InAs nanowire growth and *in-situ* InP shell passivation, shown in **Fig. 2c** and **2d**, respectively. A representative single InAs/InP core-shell nanowire (**Fig. 2e**) displays a uniform hexagonal morphology bounded by well-defined $\{11\bar{2}0\}$ sidewall facets, indicating conformal InP passivation. The nanowire arrays exhibit excellent morphological uniformity and near-unity growth yield (**Supplementary Fig. S2**) across a wide range of diameters (300-600 nm), establishing the foundation for subsequent spectral encoding through geometry engineering. As an illustration of size constraints, nanowires with diameters >300 nm, but without buffer layers yielded defective hillocks and excessive lateral growth (**Supplementary Fig. S3**). This evidences the critical role of the buffer layer in enabling diameter control of our devices.

The crystal structure of the InAs/InP core-shell nanowires was examined by aberration-corrected transmission electron microscopy (TEM). The selected-area electron diffraction (SAED) pattern taken at the middle part of a single nanowire, shown in **Fig. 2f**, reveals diffraction spots associated with both wurtzite (WZ) and zinc-blende (ZB) phase crystal planes, together with stacking fault-induced streak features along the WZ [0001]/ZB [111] axial direction. To further elucidate the crystal structure, we performed high-angle annular dark-field scanning transmission electron microscopy (HAADF-STEM) on a single nanowire. As shown in **Fig. 2g**, the nanowire exhibits a dominant WZ crystal phase with a high density of stacking faults along the axial growth direction, corroborating the streak features shown in the SAED pattern. The WZ-dominant InAs crystal structure is favourable for room-temperature photodetection because the slightly wider bandgap of WZ InAs (0.38 eV) relative to ZB InAs (0.36 eV) reduces the thermally generated bulk carrier population that contributes to leakage current.[32] The corresponding energy-dispersive X-ray spectroscopy maps show an arsenic-rich interior and enhanced phosphorus intensity at the nanowire periphery, while indium remains distributed across both regions, consistent with an InAs core surrounded by a ~ 15 nm InP rich shell (**Fig. 2h**). A weak phosphorus like background is also observed within the nanowire core, which might be caused by an overlap between phosphorus Kα and platinum Kα lines from the focused ion beam-deposited protective platinum layer. Overall, the morphologically uniform and WZ-dominant highly crystalline InAs/InP core-shell nanowire array growth provide the critical materials basis for the room-temperature e-SWIR devices described below.

**Device design and optimization for self-powered room-temperature e-SWIR photodetection and single-pixel imaging**

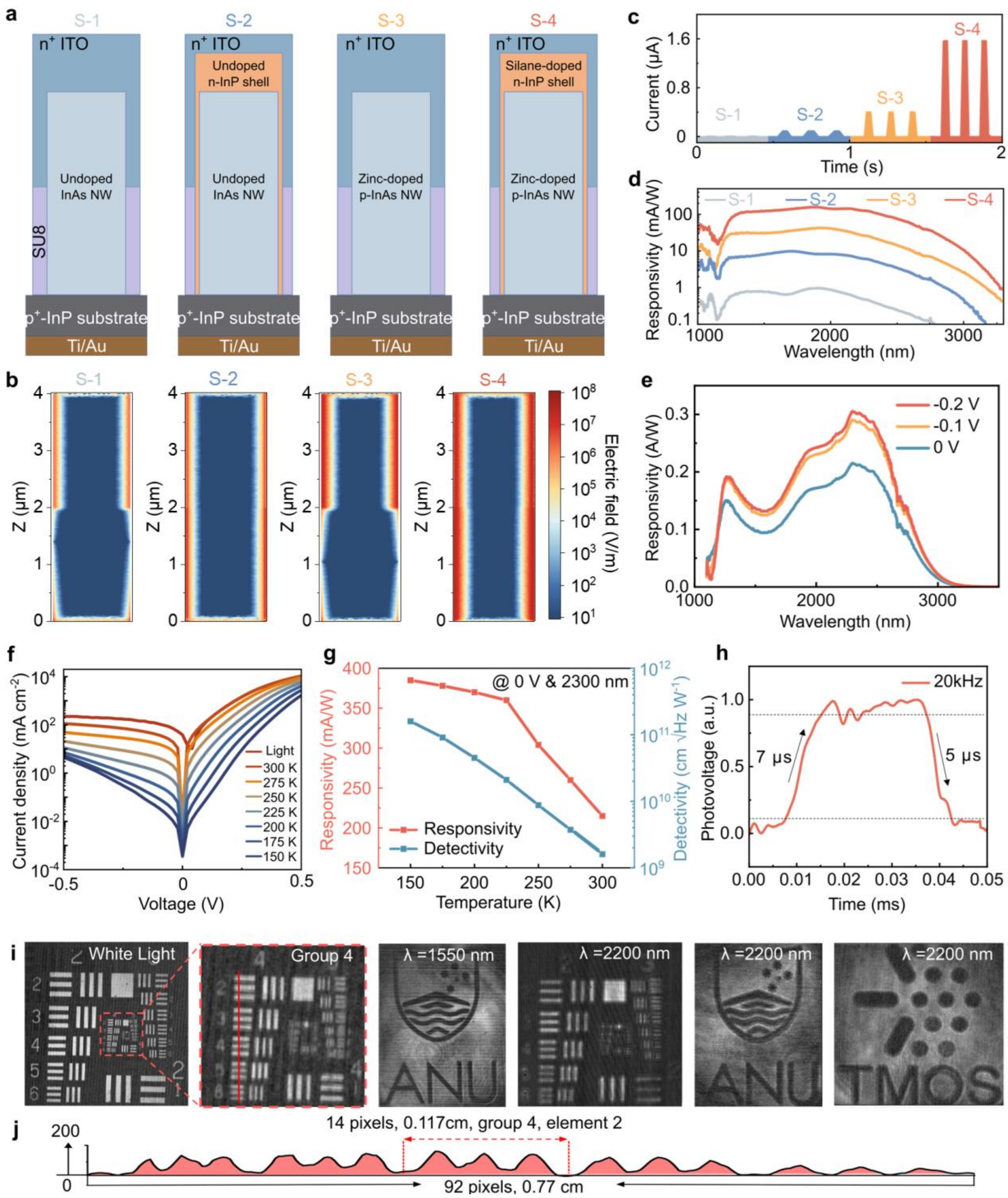

**Figure 3 | Design and optimization of nanowire photodetector structures for room-temperature e-SWIR detection. a&b**, Cross-sectional schematics (a) and simulated radial electric field distributions, |E|, within nanowire at zero bias (b) of four device structures (S1-4) with different nanowire radial junction configurations. **c&d**, Comparison of photocurrent transients (c) and responsivity spectra (d) obtained from the four device structures shown in

(a), showing progressive improvement from S-1 to S-4. The strongest photoresponse is obtained from the p-InAs/n-InP core-shell nanowire device with an $n^+$-ITO top contact (S-4). **e**, Bias-dependent responsivity spectra of the optimized device structure (S-4) measured at 300 K. **f**, Current density-voltage characteristics of S-4 measured under illumination and in the dark at temperatures from 150 K to 300 K. **g**, Temperature-dependent responsivity and specific detectivity measured at 2300 nm. **h**, Normalized single-pulse temporal response of the photovoltage to pulsed laser illumination (1550 nm) measured at 20 kHz with estimated rise/fall time of 7/5 μs. **i**, High-resolution single-pixel images recorded by an InAs nanowire detector under white light from an LED and with laser illumination ($\lambda = 1550$ nm and 2200 nm) for various targets. **j**, The line profile of elements 1-6 (red line) of group 4 from (**i**).

This material platform was utilized to fabricate 4 photodetector device structures with nanowire diameter of 530 nm and different core doping profiles. The radial junction architectures (**Figure 3a)** were designed and experimentally evaluated (see device fabrication details in **Methods**). Devices S-1 and S-2 employ unintentionally doped n-InAs cores without and with an unintentionally doped n-InP shell, respectively; whereas S-3 and S-4 employ zinc-doped p-InAs cores, with S-4 further incorporating a silane-doped n-InP shell. It is worth noting that although no dopant precursor was introduced during core growth in S-1 and S-2, the InAs nanowires exhibit unintentional n-type conductivity owing to background donor incorporation during MOCVD and surface electron accumulation. Simulated electric field distributions (details in **Methods**) under zero bias reveal distinct carrier-separation characteristics among the four device architectures (**Fig. 3b**). In S-1 and S-2, the field arises from both the axial n-InAs/$p^+$-InP substrate junction near the nanowire base and radial junction at the n-type nanowire/contact and core-shell interfaces. These isotype interfacial fields, however, are much less effective for electron-hole separation than the p-InAs/n-InP radial junction in S-4. In the shell-free S-1 and S-3 structures, the radial field is confined mainly to the ITO-contacted region. In contrast, the p-InAs/n-InP core-shell junction in S-4 produces a strong and continuous radial field along the nanowire sidewalls, facilitating carrier separation and extraction throughout the photoactive volume. The corresponding photocurrent transients measured under 1800 nm illumination, together with the responsivity spectra, are shown in **Fig. 3c, d**. As anticipated, among the four structures, S-4, exhibits the highest photocurrent and the strongest broadband response, benefiting from the strong built-in electric field formed by the p-InAs/n-InP radial heterojunction around the nanowire sidewalls enabling efficient photogenerated carrier separation and collection; while the InP shell suppresses surface recombination, which is known to limit the performance of As-based nanowire devices.[33,34]

As shown in **Fig. 3e,** the S-4 device exhibits a broad e-SWIR response, with a long-wavelength cut-off of ~ 3.3 μm arising from the predominantly WZ crystal structure of the selective-area-grown InAs nanowires. At 2300 nm, the peak responsivity increases from 0.215 A $W^{-1}$ at 0 V to 0.289 A $W^{-1}$ at -0.1 V and 0.306 A $W^{-1}$ at -0.2 V. The small additional increase between -0.1 and -0.2 V indicates that carrier collection approaches saturation under modest reverse bias. Based on the substantial responsivity obtained at zero bias, our subsequent measurements were performed in the self-powered operation. The photocurrent scales linearly with incident optical power (**Supplementary Fig. S4**), indicating stable operation over a broad illumination range. The measured voltage noise, current noise spectral density and detectivity spectra are shown in **Supplementary Fig. S5**. At room-temperature and zero bias, the device exhibits a low current noise density of $2.78 \times 10^{-12}$ $A \cdot Hz^{-1/2}$, corresponding to a noise-equivalent power of $1.29 \times 10^{-11}$ $W \cdot Hz^{-1/2}$ and peak detectivity of $1.6 \times 10^{9}$ $cm \cdot Hz^{1/2} \cdot W^{-1}$ at 2300 nm, which is comparable to that reported in state-of-the-art commercial photodetectors (approximately $3 \times 10^{9}$ $cm \cdot Hz^{1/2} \cdot W^{-1}$).[18]

The temperature-dependent current density-voltage (*J*-*V*) characteristics are shown in **Fig. 3f**. At room-temperature, the *J*-*V* characteristics demonstrates good diode behaviour with a rectification ratio of 125, confirming a well-formed radial p-n heterojunction between the p-InAs/InP nanowire and the electron selective contact of ITO. The dark current density *J* of our device is lower than those of some of the best-performing InAs p-i-n photodetectors reported to date (**Supplementary Fig. S6),**[18,19,25,35] including photovoltaic detectors,[18] avalanche photodetectors,[35] and photodiodes,[19] under the same operating bias of -0.02 V at 300 K. Notably, the dark current density of our devices at -0.02 V (10.9 $mA/cm^2$) is lower than that of the reported InAs detectors, owing to the significantly reduced material volume and effective surface passivation in our nanowire-based detectors. As the temperature gradually decreases to 150 K, the dark current density is further reduced by more than three orders of magnitude compared to that at room-temperature. Consistent with the suppressed dark current at lower temperatures, both the responsivity and detectivity at 2300 nm increase upon cooling, reaching 0.38 A $W^{-1}$ and $1.81 \times 10^{11}$ $cm \cdot Hz^{1/2} \cdot W^{-1}$, respectively, at 150 K, as shown in **Fig. 3g**. The device also shows fast temporal response, with a rise time of 7 μs and fall time of 5 μs under a modulated illumination at 1550 nm, respectively (**Fig. 3h**), which are 1-2 orders of magnitude faster than e-SWIR photodetectors reported in the literature.[24,36-40] The 3-dB cut-off frequency reaches 60 kHz (**Supplementary Fig. S7**), indicating the capability for high-speed infrared sensing.

We integrated the InAs nanowire array photodetector into a DMD-based single-pixel imaging system to further demonstrate the imaging capability of the device. A schematic of the setup and experimental details are shown in **Supplementary Fig. S8** and **Methods**. The fast photoresponse of our device, with rise and fall times of ~7/5 μs, enables rapid detection of successive DMD patterns and drastically reduces the acquisition time in single-pixel imaging.[41] Using a single InAs nanowire array device, images of multiple targets were reconstructed under white-light, 1550 nm, and 2200 nm illumination (**Fig. 3i**). Under white-light illumination, the reconstruction resolves Group 4, Element 5 of the 1951 USAF resolution chart, corresponding to a spatial resolution of 25.39 line pairs per millimetre (lp/mm) (**Fig. 3j**). In single-pixel imaging, the spatial resolution is mainly determined by the detector active area, DMD pixel size, and numerical aperture of the imaging optics.[41] Using the same imaging configuration, a high-resolution ANU logo image of 305 × 236 pixels with clear edges was further reconstructed under 1550 nm illumination. Under 2200 nm illumination, clear images of the USAF 1951 target, ANU logo, and TMOS logo were also obtained, demonstrating the e-SWIR imaging capability of the device. A weak non-uniform background is observed due to spatial variation of the illumination beam. The reconstructed images demonstrate clear contrast and target features, highlighting the broadband sensitivity, fast response, and self-powered room-temperature imaging capability of the InAs nanowire photodetector providing significant improvement in size, weight, power and cost requirements.

## Spectral reconstruction with a compact nanowire-array e-SWIR spectrometer

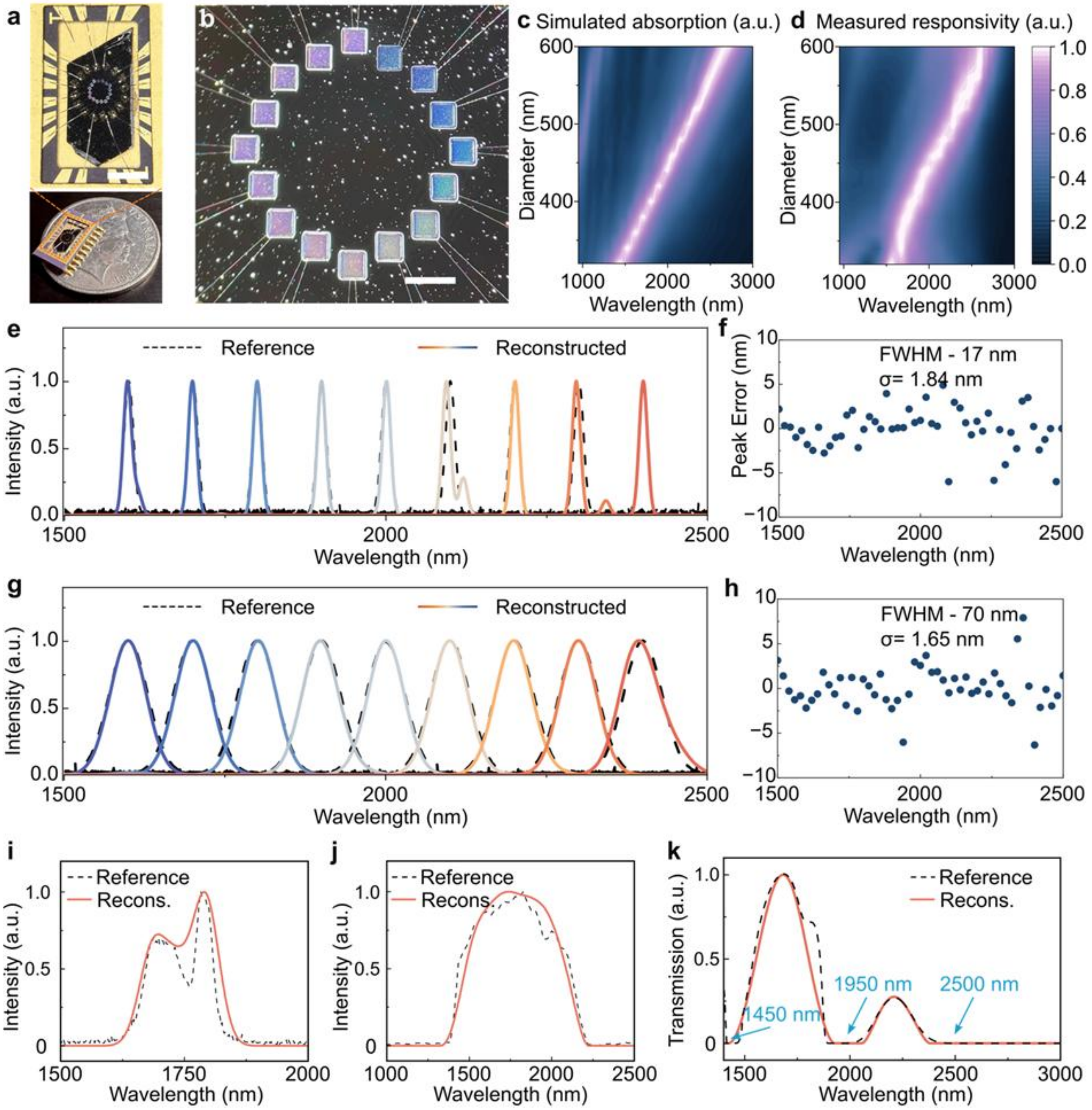

**Figure 4 |Spectral reconstruction with a compact InAs nanowire-array e-SWIR spectrometer.** **a**, Optical image of the wire-bonded spectrometer chip, shown alongside an Australian 50-cent coin for size comparison. **b**, Magnified optical image of the 16-pixel detector array. Scale bar, 200 μm. **c**, **d**, Simulated absorption (**c**) and measured normalized responsivity (**d**) maps of the nanowire-array pixels, showing geometry-dependent spectral shifts across the e-SWIR. **e, g**, Reconstructed narrowband single-peak spectra and measured reference spectra for monochromatic inputs with bandwidth of 17 nm (**e**) and 70 nm (**g**). **f, h**, Mean peak-localization error between reconstructed and reference spectra from (**e**) and (**g**). **i**, Reconstruction of a dual-peak spectrum with unequal intensities and a peak separation of 93

nm. **j**, Reconstruction of a broadband spectrum with a bandwidth exceeding 700 nm. **k**, Reconstructed water-transmission spectrum compared with the FTIR reference. Characteristic absorption features near 1450, 1950 and 2500 nm are identified.

Based on the optimized heterojunction device structure, we further vary the nanowire array parameters, i.e., height, diameter and pitch, to generate spectrally distinct responses required for computational spectrometry. FDTD simulations were performed to investigate the geometry-dependent absorption of the InAs nanowire arrays. Periodic arrays of subwavelength nanowires can support Mie-type and waveguide-like optical resonances,[20,22,23,42-44] enabling both enhanced absorption and geometry-controlled spectral tunability. The simulations show that nanowire height primarily determines the overall absorption strength, whereas the array pitch controls the modal composition and spectral bandwidth (**Supplementary Fig. S9**). At larger pitches, weaker coupling between the nanowires suppresses secondary modal contributions and produces narrower, more spectrally isolated resonances[11] which enhance wavelength discrimination and spectral resolution. These optical design principles guided the selection of nanowire array geometry for constructing the multipixel response matrix used in the computational spectrometer.

The geometry-dependent optical resonances were translated into a functional computational spectrometer by integrating 16 nanowire-array detectors into a compact wire-bonded chip through the aforementioned epitaxy growth and device fabrication processes (**Figure 4a, b**), with nanowire diameters of 300 - 600 nm in 20 nm increments. SEM images, dark current–voltage characteristics, and normalized photocurrent response matrix of the individual pixels are presented in **Supplementary Figs. S10-11**, confirming the morphological and electrical consistency of the detector library corresponding to a set of distinct wavelength-dependent responses measured from 15 nanowire-array pixels in e-SWIR used for spectral reconstruction. Note that a non-functional pixel was excluded in the figure. In summary, the measured responsivity map reproduces the geometry-dependent spectral evolution predicted by the simulated absorption map (**Fig. 4c, d**), including the red shift of the dominant response band from ~ 1500 nm to 2700 nm. This agreement between the simulated and measured results show that the designed optical resonances are preserved after material growth and device fabrication, demonstrating the practical applicability of the nanowire arrays as spectral encoding elements that serve the dual functions of detection and external filtering within the aperture.

Incident spectra were reconstructed from the measured photocurrent vector using the calibrated responsivity matrix through a regularized inverse framework, following the general principle used in computational spectrometers.[8] In the discretized form, the measured photocurrent vector can be written as $I = RP_{\lambda}+n$, where $R$ is the calibrated responsivity matrix, $P_{\lambda}$ is the incident spectrum, and $n$ is measurement noise. To improve reconstruction stability with a limited number of detector channels, the incident spectrum was represented using a non-negative Gaussian basis and reconstructed by Tikhonov-regularized non-negative least-squares fitting. For monochromatic narrowband inputs, this Gaussian basis representation is consistent with the expected single-peak spectral profile. Details of the reconstruction algorithm are provided in the **Methods**. The spectral reconstruction capability was first evaluated using single-peak incident spectra with full-width at half-maximum values of 17 and 70 nm, as shown in **Fig. 4e and g**. The mean peak-localization error, calculated as the mean absolute difference between the reconstructed and reference peak positions, is ~ 1.84 nm and ~1.65 nm across the measured spectral range, demonstrating accurate peak localization for single-peak spectra (**Fig. 4f and h**). Beyond single-peak spectra, the spectrometer can also resolve 2 narrowband peaks with a spacing of 93 nm and unequal intensities (**Fig. 4i**), and reconstruct broadband spectra with bandwidths >700 nm (**Fig. 4j**) using broader Gaussian basis functions in the reconstruction framework. The measurement setup is shown in **Supplementary Fig. S12.** We further assessed the potential application for molecular spectroscopy by reconstructing the transmission spectrum of water using the infrared source of a Fourier-transform infrared (FTIR) spectroscopy system, shown in **Supplementary Fig. S13**. As shown in **Fig. 4k**, 3 characteristic water absorption features centred at ~ 1450 nm, 1950 nm, and 2500 nm are clearly resolved, demonstrating good agreement with the reference FTIR spectrum. Our device demonstrates the better single-peak resolving capability of 17 nm, together with a dual-peak resolving capability of 93 nm compared to other e-SWIR computational spectrometers (**Supplementary Tables S2**).[6,7,17] Moreover, to the best of our knowledge, this is the first reported computational spectrometer that enables self-powered room-temperature operation with minimal energy consumption.

## Application for e-SWIR hyperspectral imaging

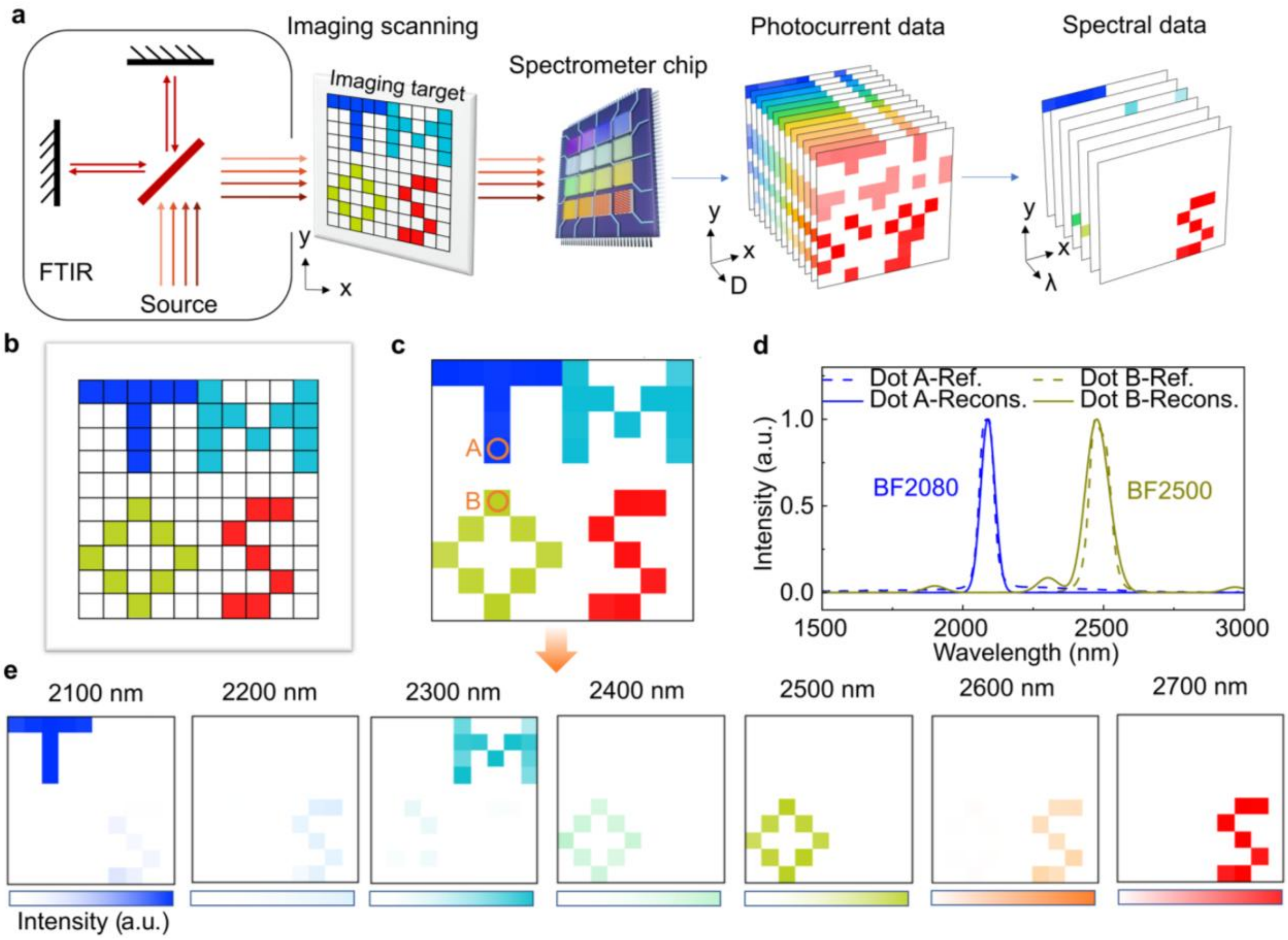

**Figure 5 | Demonstration of e-SWIR hyperspectral imaging. a**, Schematic illustration of the spectral imaging workflow. The modulated spectrum from the FTIR source is projected onto the imaging target, detected by the InAs nanowire array spectrometer chip, and reconstructed into spectral–spatial data cubes. **b**, Original target pattern for spectral imaging consisting of bandpass filters (BF) of 2080 (T), 2290 (M), 2500 (O), 2700 nm (S) and blocking areas. **c**, Reconstructed spectral imaging (10×10 pixels) obtained with the InAs nanowire spectrometer chip. **d**, Spectral reconstruction results at positions A and B in c, compared with the corresponding reference spectra measured by FTIR. **e**, Reconstructed images at 2100, 2200, 2300, 2400, 2500, 2600, 2700 nm, extracted from the spectral imaging dataset.

The same spectrometer chip can also be employed for hyperspectral imaging, which combines the spatial resolution of conventional imaging with the spectral resolution of spectroscopy.[3,4,10,45] As illustrated in **Figure 5a**, an FTIR source was projected through a spectrally encoded shadow mask and then focused onto the InAs nanowire spectrometer chip. The mask consisted of a 10 × 10 pixelated patterns, in which the letter-shaped apertures spelling "TMOS" were defined in an opaque hard mask and spectrally encoded using different bandpass filters (BF) placed behind the corresponding letters (**Fig. 5b**). The photocurrent responses from the nanowire array pixels at different spatial positions were collected to construct a complete photocurrent response matrix. Spectral reconstruction was then performed at each pixel to generate a three-dimensional spectral data cube containing spatial (X, Y) and spectral (λ) information (details see **Supplementary Fig. S14** and **S15**).

The reconstructed image shown in **Fig. 5c** was obtained by assigning false colours to the reconstructed spectral features, reproducing the original spectrally encoded "TMOS" target in **Fig. 5b**. To further validate the spectral accuracy, the reconstructed spectra at two representative positions, A and B, were compared with the corresponding reference spectra in **Fig. 5d**, showing good agreement with the bandpass-filter-defined spectral profiles. Representative wavelength slices extracted from the reconstructed cube are shown in **Fig. 5e**. Owing to the narrow bandwidth of the BF2080 filter, the blue letter "T" appears predominantly in the 2100 nm channel. In contrast, the green letter "O" is visible across the 2300 nm, 2400 nm and 2500 nm channels, reflecting the broader spectral profile of the corresponding BF2500 filter. This wavelength-dependent variation arises from the different transmission characteristics of the filters used for spectral encoding and illustrates the richer information content available from hyperspectral intensity mapping compared with a single reconstructed image. Such wavelength-resolved spatial information is valuable for compact infrared imaging applications, including molecular mapping, biomedical tissue analysis, and lesion assessment.[46]

**Supplementary Tables S1** and **S2** benchmark our InAs nanowire array radial heterojunction based photodetector and spectrometer against reported e-SWIR photodetectors and computational spectrometers, respectively. Our InAs/InP nanowire detectors have excellent performance for broadband and geometry-tunable response, responsivity, detectivity, and response times under room-temperature and zero-bias operating conditions. The findings in literature can be summarized to provide a quick comparison with the performance improvements made my our system. Previuosly reported computational spectrometers based on two-dimensional materials,[5-8] colloidal quantum dots,[2] organic photodiodes,[3,4] Si[11,17] and III-Vs[10,47] have shown impressive spectral reconstruction performance, in the visible and near-infrared ranges. However, in the e-SWIR regime, systems based on doped-Si[17] or black phosphorus[48,49] show coarser spectral resolution (on the order of tens of nanometers) with limited responsivity, detectivity, and/or stability in ambient conditions. In comparison, our chip-scale, room-temperature, self-powered InAs/InP nanowire spectrometer outperforms these reported systems in sensitivity and stability, indicating a viable path towards scalable, filter-free computational spectrometry for e-SWIR.

# Conclusion

In summary, we present a first-ever demonstration of a room-temperature computational spectrometer that can operate across the entire e-SWIR region based on InAs/InP core-shell nanowire detector arrays monolithically integrated on InP substrates. With an InAsP double-buffer layer growth strategy, we overcame the growth challenges associated with the large nanowire-substrate lattice mismatch, realizing highly-uniform InAs nanowire arrays with widely tailorable diameters on a low-noise non-native InP substrate. This deterministic diameter control enables the geometry-based spectral encoding, that forms the basis of the nanowire-based computational spectrometer that has multifunctionality of detection and filtering built into the aperture. Furthermore, through doping control and effective *in-situ* nanowire surface passivation, we engineered a self-powered radial p-InAs/n-InP heterojunction architecture for efficient carrier separation, dark current suppression, enabling high performance room-temperature e-SWIR photodetection and single-pixel imaging. By systematically engineering the nanowire array geometry, including the diameter and pitch, we modulated the wavelength-dependent photoresponse across the 1–3 μm range, to establish a tunable and spectrally diverse response matrix for computational reconstruction. Finally, through the growth and fabrication of a compact multipixel detector array chip, we successfully demonstrated filter-free spectral reconstruction with a mean peak-localization error as low as 1.65 nm, and applied it for molecular absorption sensing and hyperspectral imaging.

Future work will include steps to incorporate electrical tunability into this platform,[10,47] so that the geometry-programmed response matrix demonstrated here can be further expanded by dynamic control of carrier collection or optical response. This could reduce the number of physical detector elements required for spectral reconstruction while enabling enhanced spectral resolution and reconfigurable operation modes, leading to multifunctional e-SWIR systems that combine spectrometry, imaging, and adaptive sensing. Furthermore, integration with polarization-sensitive InAs nanostructures, such as nanosheet geometries,[50,51] could allow application of the InAs-based platform in computational imaging to simultaneously resolve spectral, spatial, and polarization information for next-generation multidimensional infrared sensing and imaging systems.

# Methods

**Substrate patterning for growth**: First, a 30 nm-thick $SiO_2$ growth mask layer was deposited on a heavily p-doped (111)B InP substrate by atomic layer deposition (ALD) at 300 °C. A negative photoresist (AR6200.04) of ~ 100 nm thickness was spin-coated onto the $SiO_2$ layer and baked at 150 °C for 1 min on a hotplate. The resist was then patterned using a Raith 150 electron beam lithography system to form 200 × 200 μm arrays with the designed mask opening and pitch sizes. After resist development, the exposed pattern was transferred into the $SiO_2$ growth mask by reactive-ion etching (RIE), followed by resist removal. To remove the damaged layers on the exposed substrate surface, a chemical trim-etching process was then performed on the patterned substrates using a diluted mixture of $H_2O_2$ and $H_3O_4$ solutions. The substrates were then immediately transferred into the MOCVD reactor for nanowire growth.

**MOCVD growth**: InAs nanowires were grown on these patterned substrates using an AIXTRON close-coupled showerhead reactor at a pressure of 100 mbar. Purified $H_2$, trimethylindium (TMIn), phosphine ($PH_3$), arsine ($AsH_3$), diethylzinc (DEZn), and silane ($SiH_4$) were used as the carrier gas, group III and group V precursors respectively. As shown in **Supplementary Fig. S1**, in the first step of the growth sequence, the growth samples were annealed at 750 °C under $PH_3$ overpressure for 10 minutes to fully remove the native oxide formed on the substrate surface. Then, the temperature was ramped down to 600 °C, and the P-rich and As-rich InAsP double-buffer layers were successively grown on InP substrate for 1 min (~ 20 nm) and 40 s (~ 10 nm), respectively. Throughout the buffer layer growth, TMIn was kept at a constant flow rate of $6 \times 10^{-6}$ mol/min, while the $PH_3$ and $AsH_3$ flow rates used for the P-rich InAsP buffer layer were $8.5 \times 10^{-3}$ and $2.6 \times 10^{-4}$ mol/min, and for the As-rich InAsP buffer layer were $1.3\times 10^{-3}$ and $7.1\times 10^{-3}$ mol/min, respectively. Next, InAs nanowire growth was performed at 690 °C for 10 min by switching off the $PH_3$ flow while maintaining the flow rates of TMIn and $AsH_3$. To suppress surface recombination, the nanowires were passivated by growing a thin layer of InP at 600 °C for 20 s.

**Material characterization**: The morphology of the InAs nanowire arrays was characterized using a FEI Verios 460 field-emission SEM. Their crystal structure and elemental distribution were examined using an aberration-corrected FEI Themis Z300 TEM equipped for HAADF-STEM and energy-dispersive X-ray spectroscopy. Site-specific TEM lamellae containing individual nanowires were prepared using focused ion beam lift-out. A protective Pt layer was deposited before milling, followed by progressive thinning and final low-voltage polishing to obtain electron-transparent lamellae while minimizing ion-beam-induced damage.

**Device fabrication**: Firstly, the InAs nanowire array was spin-coated with a layer of SU8 photoresist. Then, inductively coupled plasma (ICP) etching using $CHF_3$ plasma was carried out to expose the nanowire tips. This was followed by a UV exposure step to cure the SU8 film. Finally, a 100 nm-thick ITO layer was deposited on top of the nanowire array as the top contact by sputter deposition and 10 nm Ti and 100 nm Au were deposited at the back of the substrate as the back contact by electron-beam evaporation.

**Device electrical simulation:** The energy band diagram and built-in electric field profiles at thermal equilibrium were simulated using the CHARGE solver in the Ansys Lumerical software suite, which solves a finite-element method based semiconductor drift-diffusion model. A two-dimensional y-normal x-z cross-section of a nanowire with 530 nm diameter and 4 µm height grown on a $p^+$-InP substrate was used. The InAsP buffer layers and lithographic mask opening were neglected to simplify the model. Four device configurations, from S-1 toS-4 as shown in Fig.3a, were simulated: S-1 and S-2 contained unintentionally doped n-InAs cores ($1\times10^{16}$ $cm^{-3}$), whereas S-3 and S-4 contained zinc-doped p-InAs cores ($1\times10^{18}$ $cm^{-3}$). S-2 incorporated an unintentionally doped n-InP shell ($1\times10^{16}$ $cm^{-3}$), while S-4 incorporated a silane-doped n-InP shell ($1\times10^{18}$ $cm^{-3}$). S-1 and S-3 were simulated without an InP shell. The $p^+$-InP substrate doping concentration was set to $5\times10^{18}$ $cm^{-3}$. The InP shell thickness was set to 15 nm. A conformal 100 nm thick ITO electrode contacted the exposed upper sidewalls and top of the nanowire above the 2 µm planarization level. The bottom Au contact was treated as an Ohmic contact, while the top ITO contact was directly modelled as a heavily n-doped semiconductor. Both contacts were maintained at 0 V to obtain the equilibrium built-in electric-field distribution and energy band diagrams. The radial electric-field magnitude, $|E|$, was extracted for comparison among the four structures, as shown in **Fig.3b**. The calculated energy band diagram of the candidate device structure (S-4) at zero bias is shown in **Fig.1g**. For S-4 with p-type InAs nanowires as the light absorber, photogenerated electrons are the minority carriers contributing to the photocurrent, which can easily tunnel through the thin InP barrier and be collected by the electron-selective contact (ITO). In contrast, for other structures with an n-type InAs core (band diagram not shown), photogenerated holes are the minority carriers, and their extraction is blocked by the large valence band offset at the InAs/ITO interface.

**FDTD optical simulation**: Three-dimensional simulations were performed using Ansys Lumerical FDTD Solutions. A periodic unit cell representing the nanowire array was modelled with periodic lateral boundaries and perfectly matched layers along the nanowire axis. A normally incident broadband plane wave covering 0.6–3.5 μm was used, and absorption was calculated as $A=1-R-T$. With the nanowire diameter fixed at 500 nm, the height and pitch were sequentially varied; a height of 3.5 μm was selected for the pitch sweep. At representative absorption peaks, the fields were projected onto the first ten eigenmodes using a mode-expansion monitor. The dominant modes were assigned as $HE_{21}$-like and $HE_{11}$-like based on their field distributions.

**Photodetector characterization**: The *I–V* characteristics of the nanowire array photodetectors were measured using a KEYSIGHT B2902A Precision Source/ Measure Unit. The photocurrent spectra of the detectors were measured using the conventional amplitude modulation technique with a tungsten–halogen lamp as a white illumination source, a mechanical chopper (frequency 333 Hz), an Acton SpectraPro® 2300i monochromator, a Stanford SR570 low-noise current preamplifier, and a Stanford SR830 DSP lock-in amplifier. A FTIR Bruker VERTEX 80v spectrometer (calibrated using a commercial PbSe detector) and a Stanford SR570 preamplifier were used to measure the photoresponse of the e-SWIR nanowire array detectors. The temporal response of the nanowire array detectors was measured by laser illumination at a wavelength of 1550 nm modulated at different frequencies, with the photoresponse signal amplified using an SR570 low-noise current preamplifier and recorded using a LeCroy Waveace 102 digital oscilloscope. All the tests were conducted at room temperature (~300 K) unless specified otherwise. The commercial spectrometer is an OceanOptics NIRQuest Spectrometer.

**Single-pixel imaging**: The imaging process involves illuminating the targets using white LED and SWIR laser sources at wavelengths of 1550 and 2200 nm. The transmitted light was focused through a lens onto a spatial light modulator (a Texas Instruments Digital Micromirror Device, DMD). Light reflected from the "on" pixels of the DMD was further focused by a second lens onto the back focal plane of an objective lens, whereas the photodetector was positioned at the focal point of the objective lens. The signal from the photodetector was processed using a trans-impedance amplifier and sampled by an analog-to-digital converter.

**Spectral reconstruction methodology**: The reconstruction process is based on the linear relationship between the measured photocurrent vector *I*, the unknown input spectrum $P_\lambda$, the calibrated responsivity matrix $R$, and the measured noise $n$, written as $I = RP_{\lambda}+n$. In our approach, the responsivity matrix was first obtained from calibration measurements, and the spectrum of the unknown incident light was then reconstructed from the measured response intensities using regularized inversion. Specifically, we applied Tikhonov regularization to stabilize the solution against noise and matrix ill-conditioning, yielding the reconstructed spectrum from the measured photocurrent signals.

**Hyperspectral imaging demonstration**: The measurement of the transmission spectrum of water was performed using the globar source of the FTIR system. The water sample was filled into a cuvette. Light transmission spectrum through the cuvette was measured and reconstructed using our miniaturized spectrometer. For the demonstration of hyperspectral imaging, a shadow mask with the transparent region colored by bandpass filters was put in front of the miniaturized spectrometer. The incident light spectrum was measured and reconstructed for each pixel.

## Data availability

All relevant data that support the findings of this work are available from the corresponding author upon reasonable request.

## Code availability

All the codes used in this paper are available from the corresponding authors upon reasonable request.

## Acknowledgements

The authors acknowledge Prof. Zhipei Sun (Aalto University) for publicly sharing the spectral reconstruction code[8], which served as the basis for the reconstruction algorithm adapted in this work. The authors acknowledge the financial support from the Australian Research Council. C.C. acknowledges the ARC Discovery Early Career Research Award (DE250100406). C.J. and L.F. acknowledge the US Air Force AOARD R&D grant (FA2386-24-1-4077), The authors also acknowledge the ACT node of the NCRIS-enabled Australian National Fabrication Facility (ANFF) at the Australian National University (ANFF-ACT) for access to epitaxy and nanofabrication facilities. Authors J.W.A and M.S.A would like to acknowledge funding for CLAWS provided by the Office of Under Secretary of Defense for Research and Engineering, Applied Research for the Advancement of S&T Priorities (ARAP) Program.

## Author contributions

Z.Y.L. M.S.A., J.W.A., and L.F. conceived and designed the research. C.C., Z.Y.L., and L.F. supervised the project. Y.Y. and W.W. carried out the material growth. Y.Y. and Z.Y.L. fabricated the devices. Y.Y., Z.Y.L., and Y.B. performed the FTIR and temperature-dependent measurements. C.C., Y.Y., and D.L. carried out the spectral reconstruction and hyperspectral imaging. J.C., S.K., and Y.Y. conducted the single-pixel imaging experiments. L.L., H.L., and X.X. performed the TEM characterization. Z.L., W.W., Y.Y., and K.D. carried out the numerical simulations. M.S.A., J.W.A., H.T., C.J., and K.C. provided valuable suggestions and discussions. Y.Y., C.C., Z.Y.L., and L.F. wrote the manuscript. All authors discussed the results and contributed to revising the manuscript.

## Competing interests

The authors declare no competing interests.